\documentclass[12pt,a4paper]{article}
\usepackage{amssymb,amsfonts,amsmath}

\title{Quantum(-like) decision making: on validity of the Aumann theorem}

\author{Andrei Khrennikov and Irina Basieva\\
International Center for Mathematical Modeling\\
 in Physics and Cognitive Sciences \\
 Linnaeus University,  V\"axj\"o-Kalmar, Sweden\\
Prokhorov General Physics Institute\\ 
Russian Academy of Science\\ Moscow, Russia
}

\begin{document}

\maketitle              

\begin{abstract} 
Through set-theoretic formalization of the notion of common knowledge, 
Aumann  proved that if two agents have the common priors, and their posteriors for 
a given event are common knowledge, then their posteriors must be equal. In this paper we
investigate the problem of validity of this theorem in the framework of quantum(-like) 
decision making.
\end{abstract}

\section{Introduction}

We remark that during recent years the mathematical formalism of quantum mechanics was widely applied to problems 
of decision making and more generally modeling of cognition,  see, e.g., the monographs \cite{40}--\cite{33} 
as well as the series of articles \cite{37}-\cite{36a}
.  This project is based on the {\it quantum-like paradigm} \cite{44}: 
that information processing by complex cognitive systems (including social systems) taking into account 
contextual dependence of information and probabilistic reasoning can be mathematically 
described by quantum information and probability theories. 

One can find evidences  of violation of laws of classical probability theory, e.g., in violation of  
the {\it law of total probability.}
Its violation have been found in various sets of statistical data, see, e.g.,
 \cite{27}, \cite{28}, \cite{19},\cite{21},\cite{22}, 
\cite{50},  \cite{44},  
\cite{32}, \cite{33}, \cite{4}. The derivation of this law
is based on the additivity of classical probability measures and the classical definition of conditional probabilities based 
on the {\it Bayes formula.} Thus the law of total probability can be violated as the result of violation of either additivity of classical
probability, cf. with Feynman's viewpoint \cite{Feynman}, or classical Bayesian rule or both jointly. One can say that this is an integral 
statistical test of classicality of probability combining its two basic features, additivity and Bayesianity. It is interesting 
to find cognitive phenomena in which just one of these factors is responsible for deviation 
from the classical probabilistic predictions.

The role of the Bayesian updating in decision making was analyzed in \cite{8} with application to 
the problem of {\it human probability judgment errors}; in \cite{36a} this analysis was performed 
for such an important psychological phenomenon as 
{\it cognitive dissonance.}  In both studies it was shown that by using quantum probability updating one can 
present consistent mathematical descriptions of aforementioned problems.

In this paper we show that the quantum generalization of the Bayesian updating leads to violation 
of the celebrating {\it Aumann theorem} \cite{Aumann}, \cite{Aumann1} which  states that {\it if two agents have the common priors,
 and their posteriors for 
a given event $E$ are common knowledge\footnote{For readers' convenience, we now present the original Aumann's 
definition of common knowledge for two agents: ``An event $E$ is common knowledge at the state of the world
$\omega$ if 1 knows $E,$ 2 knows $E,$ 1 knows 2 knows $E,$ 2 knows 1 knows $E,$ and so on.''   The aforementioned heuristic notion of common knowledge 
can be formally described by various mathematical models. 
The classical probabilistic formalization of 
this notion was presented, e.g., in \cite{Aumann}, \cite{Aumann1}. (In fact, the problem of common knowledge plays very important role
in cognitive science, psychology, philosophy,  decision making, economics. There were published numerous papers enlightening various 
aspects of common knowledge studies. We are not able to review such studies in this paper, see, e.g., easily approachable work 
\cite{Vanderschraaf} for extended bibliography.) In this paper we present a novel 
formalization of the heuristic notion of common knowledge, namely, based on quantum probability and quantum logics. However, we do not change 
the cognitive meaning of this notion. Our quantum(-like) model just describe some features of common knowledge which were 
known by experts in aforementioned areas, but were not covered by the classical probability model.}, then their posteriors must be equal; agents with the same priors 
cannot agree to disagree.} In this note we show that in {\it some contexts} agents using quantum(-like) information processing can
{\it agree to disagree} even if  they  have the common priors, and their posteriors for 
a given event $E$ are common knowledge. The most interesting problem is to find elements of classical Aumann's model 
for common knowledge whose quantum generalization induces  violation of his theorem, we preset one of sufficient conditions 
of validity of the Aumann theorem even for agents whose information processing is described by quantum information theory and 
quantum probability.

We remark that violations of the Aumann theorem in real situations were widely discussed in literature
(see, e.g., \cite{Aumann},  \cite{Vanderschraaf} for discussion). Typically such violations are related to violation 
of one of the basic assumptions of the Aumann theorem, on the common prior probability and common knowledge about 
the posterior probabilities: either the agents do not have such a prior probability or the posterior probabilities 
are not common knowledge. However, sometimes agents agree on disagree even having the common prior and common knowledge.
One of the important sources of such violations is the presence of biases contributing to irrational update of probabilities,
see appendix 1 on the discussion; in particular, about the agreement on disagree for agents proceeding with the
{\it Self Sampling Assumption.}\footnote{We also remark that a deep connection between biased decision making and quantum modeling of cognition was established 
in the framework of theory of open quantum systems, where biases were modeled as components of the ``mental environment''
\cite{6}, \cite{7}.} 

We point out that the Aumann argument is based on usage of classical Boolean logic and quantum violation of his theorem can 
be interpreted as a consequence of using of nonclassical logic, i.e., consideration of agents processing information by using a 
nonclassical logical system, see appendix 2 for a discussion. This appendix also contains a brief discussion on the role of 
the ontic and epistemic descriptions in the (anti-)Aumann argumentaion.

A brief introductions to the classical approach to the problem of agreement on disagree is presented in footnote 3 and appendix 3.

\section{Quantum(-like) approach to common knowledge}

Following von Neumann \cite{VN} and Birkhoff and von Neumann \cite{BI} 
we represent {\it events, propositions,} 
as orthogonal projectors in complex Hilbert space
$H.$ Denote the scalar product in $H$ as $\langle\cdot \vert \cdot \rangle.$ 
For an orthogonal projector $P,$ 
we set $H_P= P(H),$ its image, and vice versa, for subspace $L$ of $H,$
the corresponding orthogonal projector is denoted by the symbol $P_L.$ 
 
The set of orthogonal projectors is a {\it lattice} with the order structure:
$P\leq Q$ iff $H_P \subset H_Q$ or equivalently, for any $\psi \in H, \; 
\langle \psi \vert P \psi  \rangle \leq \langle \psi \vert Q\psi  \rangle.$ 
For a pure state $\vert \psi\rangle,$ 
we set $P_\psi= \vert \psi\rangle \langle \psi\vert,$ the orthogonal projector 
on this vector, $P_\psi \phi=  \langle \phi \vert \psi  \rangle \psi .$

Aumann's considerations \cite{Aumann}, \cite{Aumann1} are applicable for a finite number of {\it agents}, call them $i=1,2,..., N.$
These individuals are about to learn the answers to various multi-choice {\it questions}, 
to make observations. 

In our quantum-like model the ``{\it states of the world}'' are given by pure states.\footnote{The notion of possible worlds
is very complex and it has been discussed in hundreds of papers, in philosophy, knowledge theory, modal logics. One can think 
about states as representing Leibniz's possible worlds or Wittgenstein's possible states of affairs. Of course, by 
representing the states of world by pure quantum states and saying nothing about a possible interpretation of 
the wave function, quantum state, we proceed in the purely operational way. What quantum state interpretation does match with 
the notion of the ``possible worlds'' used in literature? Suprisingly, it seems that the many worlds interpretation matches
best, see also appendix 1. There is a similarity between the state of the world and the wave function of universe. However,
since we are not so much excited by the many worlds interpretation, we proceed in the purely operational approach. The 
information interpretation of the quantum state (A. Zeilinger, C. Brukner) seems to be the most appropriate for our purposes.
}
Questions posed by agents are mathematically described by self-adjoint operators, say $A^{(i)}.$   We state again that events 
(propositions) are identified with orthogonal projectors. 
For the state of the world $\psi,$ an event $P$ {\it occurs} (takes place with probability 1) 
if $\psi$ belongs to $H_P.$

To simplify considerations, we proceed in the case of the finite dimensional state space of the world,
$m=\rm{dim} H < \infty.$  Here each self-adjoint operator can be represented as a linear combination of 
orthogonal projectors to its eigen-subspaces. In particular, the questions of agents can be expressed as
$
A^{(i)} = \sum_j a_j^{(i)}   P_j^{(i)},
$
where $(a_j^{(i)})$ are real numbers, all different eigenvalues of $A^{(i)},$ and $(P_j^{(i)})$  are the orthogonal projectors onto
the corresponding eigen-subspaces.  Here $(a_j)$  encode possible answers to the 
question of the $i$th agent.  The system of projectors ${\cal P}^{(i)} = (P_j^{(i)})$ is 
the spectral family of $A^{(i)}.$ Hence, for any agent $i,$ it is   
a ``disjoint partition of unity'':
\begin{equation}
\label{PARTY}
\bigvee_{k} P_k^{(i)}  =I, \; P_k^{(i)}\wedge P_m^{(i)} =0, k\not=m. 
\end{equation}
We remark that (\ref{PARTY})
is simply the lattice-theoretical expression of the following operator equalities: 
\begin{equation}
\label{PARTY7}
\sum_{k} P_k^{(i)}  =I, \; P_k^{(i)} P_m^{(i)} =0, k\not=m. 
\end{equation}

This spectral family can be considered as 
 information representation of the world by the $i$th agent.
In particular, ``getting the answer $a_j^{(i)}$'' is the event which is mathematically described 
by the projector $P_j^{(i)}.$

If the state of the world is represented by $\psi$ and, for some $k_0,$  $P_\psi \leq P_{k_0}^{(i)},$ then
$$
p_\psi(P_{k_0}^{(i)})= \rm{Tr} P_\psi P_{k_0}^{(i)} =1\;  \rm{and, \: for}\; k\not=k_0,\;
p_\psi(P_{k}^{(i)})= \rm{Tr} P_\psi P_{k}^{(i)} =0. 
 $$ 
Thus, in this case, the event $P_{k_0}^{(i)}$ happens with the probability one and other events from  
information representation of the world by the $i$th agent have zero probability.

However, opposite to the classical case, in general $\psi$ need not belong to any concrete
subspace $H_{P_{k}^{(i)}}.$ Nevertheless, for any pure state $\psi$, there exists the minimal 
projector $Q_\psi^{(i)}$ of the form $\sum_{m} P_{j_m}^{(i)}$ such that $P_\psi \leq Q_\psi^{(i)}.$ 
Set $O_\psi^{(i)}=\{j: P_j^{(i)}\psi\not=0\}.$ Then
$Q_\psi^{(i)}= \sum_{j \in O_\psi^{(i)}} P_j^{(i)}.$  
The  projector $Q_\psi^{(i)}$ represents the $i$th agent's knowledge  about the $\psi$-world. 
We remark that $p_\psi(Q_\psi^{(i)})=1.$

Consider the system of projectors $\tilde{\cal P}^{(i)}$ consisting of sums of the projectors from ${\cal P}^{(i)}:$
\begin{equation}
\label{ha_TT0}
\tilde{\cal P}^{(i)} =\{P= \sum_m P^{(i)}_{j_m} \}. 
\end{equation}
Then 
\begin{equation}
\label{ha_TT}
Q_\psi^{(i)} = \min\{P \in \tilde{\cal P}^{(i)}: P_\psi \leq P\}. 
\end{equation}
(We remark that, since we proceed with finite dimensional Hilbert spaces, this minimum is uniquely determined.)

\medskip

{\bf Definition 1.} {\it 
For the $\psi$-state of the world and  the event $E,$ the $i$th agent knowns $E$ if}   
\begin{equation}
\label{ha}
Q_\psi^{(i)} \leq E.    
\end{equation}

\medskip

It is evident that if, for the state of the world $\psi,$ the $i$th agent knows $E,$ then $\psi \in H_E.$ In general
the latter does not imply that $E$ is known (for the state $\psi).$ However, if $\psi \in E=P^{(i)}_j,$ then 
this event is known for $i.$   The same is valid for any event of the form $E=P^{(i)}_{j_1} \vee ... \vee P^{(i)}_{j_k}
(= P^{(i)}_{j_1}+ ...+ P^{(i)}_{j_k});$ if $\psi \in H_E,$ then such $E$ is known for $i.$   

We remark that the straightforward analog of the classical definition would be based on condition $P_j^{(i)}\leq E$
for $P_\psi \leq P_j^{(i)},$ instead of more general condition (\ref{ha}). However, it would trivialize the class of 
possible states of the world.

We now define the {\it knowledge operator} $K_i$ which applied to any event $E,$ yields the event 
``$i$th agent knows that $E.$'' 
 
\medskip

{\bf Definition 2.} {\it $K_i E= P_{H_{K_i E}},$ where $H_{K_i E}= \{\phi:  Q_{\phi/\Vert \phi \Vert}^{(i)} \leq E\}.$}     

\medskip

{\bf Proposition 1.} {\it For any event $E,$ the set $H_{K_i E}$ is a linear subspace of $H.$} 

{\bf Proof.}
Take two vectors $\phi_1, \phi_2 \in H_{K_i E}$ and consider 
their linear combination $\phi =a_1 \phi_1 + a_2 \phi_2.$  We consider also the corresponding pure states 
$\psi_1 =\phi_1/\Vert \phi_1 \Vert, \psi_2=\phi_2/\Vert \phi_2 \Vert$ and $\psi=\phi/\Vert \phi \Vert.$
We have   $Q_{\psi_m}^{(i)} \leq E.$  
Thus $\psi_m = \sum_{j \in O_{\phi_m}^{(i)}}
P_j^{(i)} \psi_m.$ It is clear that $\phi$ can be represented in the form 
$\phi=  \sum_{j \in O_{\phi_1}^{(i)} \cup O_{\phi_1}^{(i)}}
P_j^{(i)} \psi.$  Therefore $O_{\psi}^{(i)} \subset  O_{\psi_1}^{(i)} \cup O_{\psi_2}^{(i)}$ and, hence,
$Q_{\psi}^{(i)}  \leq E.$ 

Thus definition 2 is consistent.   Formally 
the operator $K_i$ has the properties similar to the properties of the classical knowledge operator. However, the real logical 
situation is not so simple, see appendix 2. 

\medskip

Now, as in the classical case, we define:
$$
M_0 E = E, M_1 E= K_1 E \wedge...\wedge K_N E, ..., M_{n+1} E = K_1 M_n E \wedge ...\wedge K_N M_n E,...
$$
As usual, $M_1 E$  is the event ``all agents know that   $E$'' and so on.
We can rewrite this definition by using subspaces, instead of projectors:
$$
H_{M_1 E}= H_{K_1 E} \cap...\cap H_{K_N E},..., H_{M_{n+1} E}  = H_{K_1 M_n E} \cap ...\cap H_{K_N M_n E},...
$$
Now we define the {\it ``common knowledge''} operator, as mutual knowledge of all finite degrees:
$$
\kappa E = \wedge_{n=0}^\infty M_n E.
$$

As in the classical case we  have that ``Where something is common knowledge everybody knows it.'' 

\medskip

{\bf Lemma 1.} {\it If $\kappa E \not= 0,$ then, for each $i,$ it can be represented as}
\begin{equation}
\label{MMM}
\kappa E = \sum_m P^{(i)}_{j_m}.
\end{equation}

{\bf Proof.} Take any nonzero vector  $\phi \in H_{\kappa E}.$ Then it belongs to $H_{K_i M_n E}$ for any $n.$    
Thus $Q_{\phi/\Vert \phi\Vert}^{(i)} \leq M_n E$ and, hence, $Q_{\phi/\Vert \phi\Vert}^{(i)}\leq  \wedge_{n=0}^\infty M_n E
= \kappa E.$ Hence, for each $\phi \in  H_{\kappa E},$ we have $\phi = Q_{\phi/\Vert \phi\Vert}^{(i)} \phi.$ Thus 
 (\ref{MMM}) holds.

\section{Quantum(-like) viewpoint on the Aumann's theorem}

\subsection{Common prior assumption}

Suppose now that both agents assigned to possible states of the world the same quantum probability distribution
given by the density operator $\rho,$ a priori state. Thus they do not know exactly 
the real state of the world (the latter is always a pure state) and a possible 
state of the world appears for them as a mixed quantum state. A priori probability for 
possible states of the world is combined with the information pictures used by the agents and given by 
their partitions of unity. 

\subsection{Quantum probability update}

Consider some event $E.$ The agents assign to it probabilities after conditioning $\rho$ on  the answers 
to their questions (on their 
information representations of the world):
\begin{equation}
\label{MMMmi}
q_k^{(i)}= p_\rho(E \vert P_k^{(i)})= \frac{\rm{Tr} P_k^{(i)} \rho P_k^{(i)} E}{\rm{Tr} P_k^{(i)} \rho P_k^{(i)}}.
\end{equation}  
We remark that the agents can assign probabilities conditioned on the results of observations only for the 
answers $a_k^{(i)}$ such that $\rm{Tr} P_k^{(i)} \rho P_k^{(i)} >0.$  

Consider the events
\begin{equation}
\label{MMMmi2}
C_{q^{(i)}}\equiv \{q_k^{(i)}= q^{(i)} \} = \bigvee_{\{k: q_k^{(i)}= q^{(i)} \}} P_k^{(i)},
\end{equation}
$ i=1,...,N,$ and set
$$
C_{q^{(1)}...q^{(N)}}=\{q_k^{(1)}= q^{(1)},,,, q_k^{(N)}= q^{(N)} \}
=\bigwedge_i C_{q^{(i)}}.
$$ 

{\bf Remark 1.} Consider the classical Aumann model \cite{Aumann}, \cite{Aumann1}. Here 
\begin{equation}
\label{MMMmi1}
q^{(i)}(\omega)= p(E \vert P^{(i)}(\omega))= \frac{p(E \cap P^{(i)}(\omega))}{p(P^{(i)}(\omega))}
\end{equation}  
and 
$C_{q^{(i)}}\equiv \{\omega: q_k^{(i)}(\omega)= q^{(i)}\}.$ We remark that if for some $\omega_0$   the probability
$q^{(i)}(\omega_0)= q^{(i)},$ then, for any $\omega \in P^{(i)}(\omega_0),$ the probability $q^{(i)}(\omega)= q^{(i)}.$
Thus 
\begin{equation}
\label{MMMmi3}
C_{q^{(i)}} = \bigcup_{\{k: q_k^{(i)}= q^{(i)}\}} P^{(i)}_k,
\end{equation}  
cf. (\ref{MMMmi2}). 

\medskip

{\bf Remark 2.} So, the quantum definition (\ref{MMMmi2}) is a natural generalization of the classical definition 
(\ref{MMMmi3}). The main difference is that the classical definition is based on the Boolean logic and the quantum one 
on the quantum  logic. The quantum operation $\vee$ differs crucially from the classical operation $\cup.$  
To make the comparison clearer, we consider projection subspaces, instead of projectors. Then, for two subspaces, say 
$H_1$ and $H_2,$ the subspace $H_1 \vee H_2$ is not simply the set-theoretic union $H_1 \cup H_2,$ but 
$H_1 \vee H_2$ is the minimal subspaces containing  $H_1 \cup H_2.$  We emphasize that the quantum logic operation $\vee$ is a nontrivial 
generalization of the classical operation $\cup$ even in the case of orthogonal subspaces $H_1$ and $H_2.$ We also point out that 
quantum logic operations violate some basic laws of the Boolean logic, for example the law of distributivity for the operations 
$\vee$ and $\wedge$ is violated, i.e., for three projectors $P, P_1, P_2,$ in general $P \wedge (P_1 \vee P_2) \not= (P \wedge P_1 ) \vee
(P \wedge P_2).$ (Even the orthogonality of $P_1$ and $P_2$ does not help.)  

\medskip

We remark that, in fact, as a consequence of mutual orthogonality of projectors from the spectral family of any Hermitian operator,  
the event $C_{q^{(i)}}$ can be represented as 
\begin{equation}
C_{q^{(i)}}=   \sum_{\{k: q_k^{(i)}= q^{(i)}\}} P_k^{(i)}.
\end{equation}
Thus the event $C_{q^{(1)}...q^{(N)}}$ has representation:
\begin{equation}
C_{q^{(1)}...q^{(N)}} = \Big(\bigvee_{\{k: q_k^{(1)}= q^{(1)}\}} P_k^{(1)}\Big)
\wedge... \wedge \Big(\bigvee_{\{k: q_k^{(N)}= q^{(N)} \}} P_k^{(N)}\Big).
\end{equation}
By taking into account Remark 2 we know that in general:
\begin{equation}
C_{q^{(1)}...q^{(N)}} \not= \bigvee_{\{k_1: q_{k_1}^{(1)}= q^{(1)}\}}... \bigvee_{\{k_N: q_{k_N}^{(1)}= q^{(1)}\}} P_{k_1}^{(1)}\bigwedge
... \bigwedge P_{k_N}^{(N)}.
\end{equation}

\subsection{Interference prevents agreement}

Suppose that the possibility of $C_{q^{(1)}...q^{(N)}}$ becoming common knowledge is not ruled out completely, i.e.,
\begin{equation}
\label{ROUT}
p_\rho(\kappa C_{q^{(1)}...q^{(N)}}) >0.
\end{equation}  
Then the straightforward quantum generalization of the classical 
Aumann theorem \cite{Aumann}, \cite{Aumann1} would imply that $q^{(1)}=...=q^{(N)}.$ However, this is not the case! (as it may be expected, 
since the classical Aumann theorem 
was heavily based on usage of Boolean logics). 

By Lemma 1  the common knowledge projector can be represented as $\kappa E = \sum_j P^{(i)}_{k_j}, i=1,...,N.$
For each such $P_{k_j}^{(1)},.., P_{k_j}^{(N)},$ we have 
$$
p_\rho(E \vert P_{k_j}^{(1)})= q^{(1)}, ..., p_\rho(E\vert P_{k_j}^{(N)})= q^{(N)}.
$$ 
(In particular, for any such projector conditional probabilities are well defined, i.e., $\rm{Tr} P_{k_j}^{(i)} \rho P_{k_j}^{(i)} >0.$)
Consider now the conditional probability:
$$p_\rho(E\vert \kappa C_{q^{(1)}...q^{(N)}} )= \frac{\rm{Tr} \kappa C_{q^{(1)}...q^{(N)}} \rho \kappa C_{q^{(1)}...q^{(N)}} E}
{\rm{Tr} \kappa C_{q^{(1)}...q^{(N)}}\rho \kappa C_{q^{(1)}...q^{(N)}}}.$$

First we remark that, for any projector $M,$ $\rm{Tr} M\rho M= \rm{Tr} \rho M.$ Thus 
$$p_\rho(E\vert \kappa C_{q^{(1)}...q^{(N)}})= \frac{\rm{Tr} \kappa C_{q^{(1)}...q^{(N)}} \rho \kappa C_{q^{(1)}...q^{(N)}} E}{\rm{Tr} \rho \kappa C_{q^{(1)}...q^{(N)}}}.$$
By using  representation given by Lemma 1 we obtain
\begin{equation}
\label{off}
p_\rho (E\vert \kappa C_{q^{(1)}...q^{(N)}} )  = \frac{1}{\rm{Tr} \rho \kappa C_{q^{(1)}...q^{(N)}}} \Big(\sum_j \rm{Tr} 
P_{k_j}^{(i)} \rho P_{k_j}^{(i)} E + 
\sum_{j\not=m}  \rm{Tr}  P_{k_j}^{(i)} \rho P_{k_m}^{(i)} E\Big).
\end{equation}
The first (diagonal) sum can be written as 
$$
\frac{1}{\rm{Tr} \rho \kappa C_{q^{(1)}...q^{(N)}}} 
\sum_j \frac{\rm{Tr} P_{k_j}^{(i)} \rho P_{k_j}^{(i)} E}{\rm{Tr} \rho P_{k_j}^{(i)}} \rm{Tr}  \rho P_{k_j}^{(i)}=
\frac{q_i}{\rm{Tr} \rho \kappa C_{q^{(1)}...q^{(N)}}}  \rm{Tr} \sum \rho P_{k_j}^{(i)}  = q^{(i)} .
$$    
In the absence of the off-diagonal term in (\ref{off}) we get $q^{(1)}=...=q^{(N)}.$ This corresponds to the classical case. However, in general the 
off-diagonal term does not vanish -- this is {\it the interference type effect.} 

Hence, in general  {\it the Aumann theorem is not valid for ``quantum(-like) decision makers.}
Thus  {\it agents processing information in the quantum information and probability  framework 
can agree on disagree.}

Form the expression (\ref{off}) for the interference term, it is clear that it has three main contributions:
\begin{itemize} 

\item Incompatibility of information representations of agents.

\item Incompatibility of an event $E$ under consideration with  
individual information representations.

\item Incompatibility of information representations  of agents with the prior state.
\end{itemize}
\subsection{Sufficient conditions of validity of the quantum(-like) version of the Aumann theorem}

Now we present a special situation in which even quantumly thinking agents cannot agree on disagree.

\medskip

{\bf Proposition 2.} {\it Let the 
common a priori state $\rho$ is given by the unity operator normalized by the dimension. If condition (\ref{ROUT}), the assumption of common prior, 
 holds, 
then $q^{(1)}=...=q^{(N)}.$}   

\medskip

To prove this statement, we point to the fact that, for such $\rho,$ the off-diagonal term in (\ref{off})
equals to zero and the proof can be completed in the same way as in the classical case. 

If $\rho=I /\rm{dim} H,$ then all states of the world are equally possible. Thus, for all agents, the a priori state of the 
world was gained in the total absence of information about the world. In this case these agents have to come to the same posteriors 
(if their posteriors are common knowledge), though they may base their posteriors on different information: the information partitions
of unity, see (\ref{PARTY}), can be incompatible, i.e., the projectors $(P_j^{(i)})$ need not commute with the projectors $(P_j^{(s)}).$ 

We remark that, although in the rigorous mathematical framework 
a density operator cannot be scaling of the unit operator, formally one can operate with such
``generalized density operators'', see von Neumann \cite{VN}. Thus formally Proposition 2 can be 
generalized to infinite-dimensional state spaces. 

In fact, Proposition 2 is a special case of a more general statement which will be soon formulated.
However, we started with Proposition 2, since it has the very clear interpretation. The interpretation 
of the following statement is not straightforward:

\medskip

{\bf Theorem 1.} {\it Let the common a priori state $\rho$ commutes with the elements of all 
partitions, i.e., for $i=1,...,N,$ 
\begin{equation}
\label{COMt}
[\rho, P_j^{(i)}]=0
\end{equation}
for any $j.$  
If condition (\ref{ROUT}) holds, then $q^{(1)}=...=q^{(N)}.$}

\medskip

Here we again see that the interference 
term in (\ref{off}) equals to zero.

As was mentioned, the interpretation of the basic condition of this theorem is not straightforward.   
In quantum mechanics, commutativity of observables is interpreted as the condition of joint measuring.
However, commutativity of an observable and a quantum state has no direct interpretation. 
 
 \medskip
 
 {\bf Lemma 3.} {\it Let $\rho$ have non-degenerate spectrum and let the condition (\ref{COMt}) holds. Then 
the partitions $(P_j^{(1)}),...(,P_j^{(N)})$ are compatible, i.e., $[P_j^{(i)}, P_m^{(s)}]=0$ for any 
pair $j,m$ and $i,s.$ }

 \medskip
 
{\bf Proof.}  Suppose that $\rho$ has non-degenerate spectrum, i.e., $\rho= \sum_k p_k P_{e_k},$
where $p_k\not= p_m, k\not=m,$ and $(e_k)$ is the orthonormal basis consisting of eigenvectors of $\rho.$
Then, for any orthogonal projector 
$P,$ the condition $[P, \rho] =0$ implies that there exists a set of indexes $O_P$ such that 
 $P= \sum_{m\in O_P} P_{e_m}.$ This is easy to show. We have $\rho P= P\rho,$ i.e., for any 
pair of basis vectors $e_t, e_s,$ we have, on one hand,
$\langle e_t\vert \rho P \vert e_s\rangle= p_t \langle e_t\vert P \vert e_s\rangle$ and, on the other hand
$\langle e_t\vert P \rho \vert e_s\rangle =p_s  \langle e_t\vert P \vert e_s\rangle.$ Hence, for $t\not=s,$ 
$\langle e_t\vert P \vert e_s\rangle =0.$ Thus $P \vert e_s\rangle= a_s \vert e_s\rangle,$ where $a=0,1.$ Set $O_P
=\{s: a_s=1\}.$   For the projectors $P_j^{(i)},$ such sets will be denoted as $O_j^{(i)}.$  

Thus in the non-degenerate case the condition (\ref{COMt}) implies that 
$P_j^{(i)}= \sum_{k\in O_j^{(i)} } P_{e_k}.$ Then 
$P_j^{(i)} P_m^{(s)} = \sum_{k\in O_j^{(i)} \cap O_m^{(s)}} P_{e_k}=
P_m^{(s)} P_j^{(i)}.$  Thus the (quantum) information partitions 
are compatible. 

\medskip

However, if the spectrum of the state $\rho$ is degenerate,
then the condition (\ref{COMt}) does not imply compatibility of partitions 
of two agents (see Proposition 2).  

\medskip

{\bf Corollary 1.} {\it  Even in the case of incompatible (quantum) information 
partitions, it is possible  to find such common a priori (quantum) states that
it is impossible to agree on disagreeing.}

\section{Conclusion}

Agents representing and processing information in the quantum(-like) manner 
can agree on disagree. Thus our quantum(-like) model of probability update in 
the presence of common knowledge matches better with the real situation. 

Typically in classical analysis of sources of violations of the Aumann theorem (and it is often violated
in reality) the common a priori probability distribution and the presence of common knowledge are pointed 
as questionable assumptions in Aumann's argumentation. We show that the validity of these assumptions does not prevent 
from the possibility that agents agree on disagree. 

The main conclusion is that agents can simply use more general rules for processing of information and probability than 
given by the classical set-measure-theoretic model based on the Kolmogorov axiomatics of probability theory. 
And this model has its own restricted domain of applications, as any mathematical model, cf. with the Euclidean model of geometry,
and departure from it given by Lobachevsky geometry (which plays an important role in special relativity theory).
 
\section*{Acknowledgments} 

The authors would like to thank J. Acacio de Barros, H. Atmanspacher, J. Busemeyer, E. N. Dzhafarov, E. Haven, E. M.  Pothos, 
for discussions on quantum probabilistic modeling of cognitive  phenomena and especially decision making and probability update. 

\section*{Appendix 1: Biased decision making and violation of the Aumann theorem}

An important source of possible violation of the Aumann theorem is the presence of various biases in the ``heads of agents''. 
Roughly speaking any bias may destroy the purity of the Bayesian update.    
 
As a widely discussed example of the anti-Aumann bias, we consider the so called SSA-bias. A     
{\it Self Sampling Assumption} (SSA)  says you are more likely to be present in worlds where 
a greater proportion of agents which are like you,  see N. Bostrom for the detailed discussion on SSA  \cite{Bostrom}.
Except that ``agents'' can be any set of things you could have been in some sense, even if you currently
know you are not some of them. This group is called a {\it reference class.} 
Agents basing their reasoning on the SSA  and having 
different reference classes need not come to the same posterior probabilities, even if the assumptions of the Aumann theorem, about the 
common prior and common knowledge, hold true. And this is clear why. Such an agent can 
ignore some of her/his information in forming her/his reference classe, 
since it asks for the proportion of her/his reference class of whom all of her/his 
information is true. This can lead to simple ignorance of a part of information presented
in common knowledge. It is often argued that the decision making based on the selection of an appropriate reference 
class is irrational. And we agree with such evaluation of the SSA decision makers. However, we do not assign negative 
valuation to  ``irrationality'' in  the decision making. As was demonstrated in  \cite{Bostrom}, the SSA-operating is quite 
common phenomenon. Since this happens and happens in many contexts, such a behavior of agents has to be modeled mathematically.
And if in the classical probabilistic framework this is impossible (as signed in the violation of the Aumann theorem
which is heavily based on Kolmogorov probability), then it is natural to explore other probabilistic models, e.g., 
quantum probability. 
In this short note we are not able to discuss quantum modeling of SSA in more details, it will be done in one of further 
articles. We finish this discussion on SSA with the following remark on the interpretation of the wave function, quantum state.
The SSA approach to decision making matches well with the many world interpretation of the quantum state. A SSA-agent 
position her/him self as belonging to a few possible reference classes, which play here the roles of the worlds.


\section*{Appendix 2: On the logical structure of the Aumann argument}

As ia well known, he Aumann argument on the impossibility of agree on disagree 
is based on the special systems of axioms of the modal logic, the system S5.  And, of course, any deviation from this system
might lead to a violation of the classical Aumann theorem. In this paper it was shown that the usage of quantum logic can generate 
a possibility to agree on disagree. This is a good place to point out that our emphasize on similarity between the classical 
(S5) and quantum knowledge operators  is a bit provocative, since this similarity is only formal, operational, and from the 
logical viewpoint these are very different representations of knowledge. 

In fact, understanding of ``what quantum logic is from semantic viewpoint'' is a complex problem by itself, see, e.g., 
works of Garola \cite{Garola} and  of Garola and Sozzo  \cite{Garola1} and the recent paper of Khrennikov and Schumann 
 \cite{ASHU} for details. One of still debated problems is whether one can really 
assign to propositions a special ``quantum truth'' value
or it is even possible to proceed with the classical truth value. In \cite{ASHU} it was motivated that the essence of logical nonclassicality 
is the performative part of quantum mechanics and at the theoretical level one can still proceed with classical logic. Thus it was motivated 
that even in quantum physics logical nonclassicality is only due to the language representation.  Such a discussion is helpful to come 
closer to understanding the following fundamental problem: whether  quantumness is in the world or in the mind. It seems that the argument presented 
in \cite{ASHU} supports the latter, i.e., that violations of classical logic and ``quantum logical effects'' are generated by the performative 
structure used for the interpretation of some natural and mental phenomena. 

In this context it may useful to use the scientific methodology in which any scientific representation has two level, the 
ontic level and the epistemic level, see, e.g., Atmanspacher and Primas \cite{ATM}. In such an approach violation of classical logic
happens at the level of epistemic description.

\section*{Appendix 3: Classical formalization for the Aumann argument}
\label{CL}

Aumann's considerations are applicable for a finite number of {\it agents}, call them $i=1,2,..., N.$
These individuals are about to learn the answers to various multi-choice {\it questions}, 
to make observations.  

Mathematically the situation is represented with the aid of classical probability space
(based on the Kolmogorov axiomatics, 1933). Typically it is assumed that the state space 
$\Omega$ representing all possible states of the world is finite. Events 
are subsets of $\Omega.$

Each agent creates its information representation for possible states of the world based on its own
possibilities to perform measurements, ``to ask questions to the world.'' Mathematically these represetations 
are given by partitions of $\Omega: {\cal P}^{(i)}= (P_j^{(i)}),$ where $\cup_j P_j^{(i)}= \Omega$ and 
$P_j^{(i)} \cap P_k^{(i)} \emptyset, j\not=k.$ Thus an agent cannot get to know the state of the world $\omega$ 
precisely; she can only get to know to which element of its information partition $P_j^{(i)}= P_j^{(i)}(\omega)$ 
this $\omega$ belongs. The agent $i$ knows  an event $E$ in the state of the world $\omega$ if 
\begin{equation}
\label{KNOW}
P_j^{(i)}(\omega) \subset E.
\end{equation}

It is assumed that on $\Omega$ there is defined probability $p,$ {\it the common prior} of all agents. In the accordance
with the measure-theoretic model of probability theory (Kolmogorov, 1933) there is given a $\sigma$-algebra, say ${\cal F},$ 
of subsets of $\Omega,$ its elements represent events (``propositions'' in some interpretations), and there is given a probability measure 
$p$ defined on  ${\cal F}.$ In the knowledge models it is typically assumed that ${\cal F}$ is generated by agents' partitions, i.e., 
this is the minimal $\sigma$-algebra containing all systems of set ${\cal P}^{(i)}, i=1,...,N.$

 We  consider the systems of sets 
$\tilde{\cal P}^{(i)}=\{ \cup_m P^{(i)}_{j_m}\}$ consisting of  finite unions of the  elements 
of the systems ${\cal P}^{(i)}$ and the system $\tilde{\cal P} = \cap_i \tilde{\cal P}^{(i)}.$   
We recall that the {\it meet} of the partitions ${\cal P}^{(i)},$ denoted by the symbol 
 $\wedge_i {\cal P}^{(i)},$
is the {\it finest common coarsening} 
of ${\cal P}^{(i)}.$  
In particular, $\wedge_i {\cal P}^{(i)} \subset \tilde{\cal P}.$

As was proven in \cite{Aumann},
{\it an event $E$ is common knowledge at $\omega$ if $E$
contains that element of ${\cal P}^{(1)} \wedge {\cal P}^{(2)}$ (the meet) containing
$\omega.$} (See footnote 3 on the definition of common knowledge.)

This result implies that, for each $i,$ the set of all states of the world for which $E$ is common knowledge, denoted by the symbol
$\kappa E,$ can be represented (in the case $\kappa E \not= \emptyset$) in the form:
\begin{equation}
\label{ee}
\kappa E=   \cup_m P^{(i)}_{j_m}.
\end{equation}



\begin{thebibliography}{99}
 

\bibitem{40} Khrennikov, A.:  Information Dynamics in Cognitive, Psychological, Social,  and Anomalous Phenomena. 
Ser.: Fundamental Theories of Physics.  Kluwer, Dordreht (2004)

\bibitem{44} Khrennikov, A.:  Ubiquitous Quantum Structure: from Psychology to Finances. Springer, Berlin-Heidelberg-New York (2010)

\bibitem{24} Busemeyer  J. R.   and   Bruza, P. D.: Quantum Models of Cognition and Decision. Cambridge Press, Cambridge (2012)

\bibitem{33} Haven, E. and  Khrennikov, A.: Quantum Social Science. Cambridge Press, Cambridge (2013)

\bibitem{37} Khrennikov, A.: On Quantum-like Probabilistic Structure of Mental Information. 
Open Systems and Information Dynamics 11 (3), 267-275 (2004)

\bibitem{19} Busemeyer, J. R. , Wang, Z.  and Townsend, J. T.:  Quantum Dynamics of Human Decision Making. 
J. Math. Psychology 50, 220-241 (2006)

\bibitem{36}  Khrennikov, A.: Quantum-like Formalism for Cognitive Measurements. Biosystems 70, 211-233 (2003)

\bibitem{38} Khrennikov, A.: Quantum-like Brain: Interference of Minds.  BioSystems 84, 225-241 (2006)

\bibitem{27} Conte, E., Todarello,  O., Federici,  A. , Vitiello, F., Lopane,  M. , Khrennikov,  A.,   
and Zbilut, J. P.:  Some Remarks on an Experiment Suggesting Quantum-like Behavior of Cognitive Entities and Formulation of an 
Abstract Quantum Mechanical Formalism to Describe Cognitive Entity and its Dynamics. 
Chaos, Solitons and Fractals 31(5), 1076-1088 (2007)

\bibitem{21} Busemeyer, J. R., Santuy,   E., and Lambert-Mogiliansky,    E.:   Comparison of Markov and quantum models of decision making. 
In:  Bruza, P., Lawless, W.,  van Rijsbergen,   K., Sofge,  D. A., Coeke,   B., Clark,  S.  (eds.) 
Quantum interaction: Proceedings of the Second Quantum Interaction Symposium, pp.68-74.  
 College Publications, London (2008)

\bibitem{28} Conte, E.,  Khrennikov, A., Todarello,   O., Federici, A., Mendolicchio,   Zbilut, J. P.:
 Mental State follow Quantum Mechanics during Perception and Cognition of Ambiguous Figures. Open Systems and 
Information Dynamics 16, 1-17 (2009)

\bibitem{28} Conte, E.,  Khrennikov, A., Todarello,   O., Federici, A., Mendolicchio,   Zbilut, J. P.:
A Pre-liminary Experimental Verification On the Possibility of Bell Inequality 
Violation in Mental States. NeuroQuantology  6 (3),  214-221  (2008)

\bibitem{22} Busemeyer, J. R.,  Wang, Z., Lambert-Mogiliansky,  A.:  
Empirical Comparison of Markov and Quantum Models of Decision Making. 
J. Math. Psychology 53 (5), 423-433 (2009).

\bibitem{1} Acacio de Barros, J.,  Suppes,   P.: Quantum Mechanics, Interference, and the Brain. 
J. Math. Psychology  53,  306-313    (2009)

\bibitem{46} Lambert-Mogiliansky, A., Zamir,   S., and  Zwirn, H.:  Type Indeterminacy: A Model of the KT (Kahneman-Tversky)-man. 
J. Math. Psychology 53 (5), 349-361 (2009).

\bibitem{50} Pothos, E. M., and Busemeyer,  J. R.:  A Quantum Probability Explanation for Violation of Rational 
Decision Theory. Proc. Royal. Soc. B 276,  2171-2178  (2009)

\bibitem{32} Haven, E.  and  Khrennikov, A.: Quantum Mechanics and Violation of the Sure-thing Principle:
 the Use of Probability Interference and other Concepts. J. Math. Psychology 53, 378-388 (2009)

\bibitem{55} Trueblood,  J. S. , Busemeyer,  J. R.:   A Quantum Probability Account of Order Effects in Inference. 
Cognitive Science 35, 1518-1552 (2011)

\bibitem{6} Asano, M., Ohya, M., Tanaka, Y.,  Khrennikov, A., and Basieva, I.:  
On Application of Gorini-Kossakowski-Sudarshan-Lindblad Equation in Cognitive Psychology. 
Open Systems and Information Dynamics 18, 55-69 (2011)

\bibitem{7}  Asano, M., Ohya, M., Tanaka, Y.,  Khrennikov, A., and Basieva, I.: 
Dynamics of Entropy in Quantum-like Model of Decision Making.  J. Theor. Biology 281,  56-64 (2011)

\bibitem{31} Dzhafarov, E. N.  and Kujala, J. V.:  Quantum Entanglement and the 
Issue of Selective Influences in Psychology: An Overview. Lecture Notes in Computer Science 7620, 184-195 (2012) 

\bibitem{4} Aerts, D., Sozzo, S. and Tapia, J.:   A Quantum Model for the Elsberg and Machina Paradoxes. 
Lecture Notes in Computer Science 7620,  48-59 (2012)

\bibitem{12} Atmanspacher, H. and  Filk, T.: Contra Classical Causality: Violating Temporal Bell Inequalities in Mental Systems. 
J. Consciousness Studies 19(5/6), 95-116 (2012)

\bibitem{12a} Atmanspacher, H. and R\"omer, H.: Order Effects in Sequential Measurements
 of Non-commuting Psychological Observables. J. Math. Psychology 56, 274-280 (2012)

\bibitem{8}   Asano, M., Basieva, I.,  Khrennikov, A.,  Ohya, M.,  Tanaka, Y.:  Quantum-like Generalization of the
 Bayesian Updating Scheme for Objective and Subjective Mental Uncertainties. J. Math. Psychology 56,  166-175 (2012)         

\bibitem{36ad} Khrennikova, P.: Evolution of Quantum-like Modeling in Decision Making Processes. AIP Conf. Proc. 1508, 108 (2012)

\bibitem{9a}  Asano, M.,  Basieva, I.,  Khrennikov, A., Ohya, M.,  Yamato, I.:  Non-Kolmogorovian Approach to 
the Context-Dependent Systems Breaking the Classical Probability Law. Found. Phys. 43,  2083-2099 (2013)

\bibitem{36a} Khrennikova, P.:  A Quantum Framework for 'Sour grapes' in Cognitive Dissonance. 
Proceedings of Quantum Interaction-13, University of Leicester Press (2013)
 



  
\bibitem{Aumann} Aumann, R.J.:  Agreeing on Disagree. Ann. Statistics 4, 1236-1239 (1976) 

\bibitem{Aumann1}  Aumann, R. J.:  Backward Induction and Common Knowledge of Rationality. Games and Economic Behavior 8,
 6–19  (1995)


\bibitem{Vanderschraaf}  Vanderschraaf, P. and Sillari, G.;  Common Knowledge. In: The Stanford Encyclopedia of Philosophy, Zalta, E N.  (ed.),
http://plato.stanford.edu/archives/fall2013/entries/common-knowledge (2013)


\bibitem{BI} Birkhoff, J. and von Neumann, J.: The Logic of Quantum Mechanics. Annals of Mathematics 37,  823-843  (1936)


\bibitem{Feynman} Feynman,  R.  and Hibbs, A.:  Quantum Mechanics and Path Integrals. McGraw-Hill, New York  (1965)


\bibitem{VN}  Von Neuman, J.:   Mathematical Foundations of Quantum Mechanics. 
Princeton University Press, Princeton (1955)

\bibitem{Bostrom} Bostrom, N.:   Anthropic Bias: Observation Selection Effects in Science and Philosophy, Ser. Studies in Philosophy.
Routledge Publ. (2010)

\bibitem{Garola} Garola, C.: A Pragmatic Interpretation of Quantum Logic. Preprint arXiv:quant-ph/0507122v2

\bibitem{Garola1} Garola, C.  and Sozzo, S.  Recovering nonstandard logics within an extended classical framework.
 Erkenntnis 78, 399-419  (2013)

\bibitem{ASHU} Khrennikov, A. and A. Schumann, A.: Quantum Non-objectivity from Performativity of Quantum Phenomena. 
Preprint 	arXiv:1404.7077 [physics.gen-ph]. To be published in Physica Scripta

\bibitem{ATM}   Atmanspacher, H.  and Primas, H.:  Epistemic and Ontic Quantum
Realities. In:  Adenier, G.,  Khrennikov, A. (eds) Foundations
of Probability and Physics-3, Conf. Proc. Ser.  750, pp. 49-62.  AIP, Melville, NY (2005).


\end{thebibliography}
\end{document}